# DIVERSITY OF CORTICAL STATES AT NON-EQUILIBRIUM SIMULATED BY THE FERROMAGNETIC ISING MODEL UNDER METROPOLIS DYNAMICS

Luciano da Fontoura Costa
Instituto de Física de São Carlos,
Universidade de São Paulo,
Caixa Postal 369,
13560-970, São Carlos, SP, Brazil
*luciano@if.sc.usp.br*

and

Olaf Sporns
Department of Psychological and Brain Sciences
Indiana University
Bloomington, IN 47405, USA
*osporns@indiana.edu*

**ABSTRACT**

This article investigates the relationship between the interconnectivity and simulated dynamics of the thalamocortical system from the specific perspective of attempting to maximize the diversity of cortical states. This is achieved by designing the dynamics such that they favor opposing activity between adjacent regions, thus promoting dynamic diversity while avoiding widespread activation or de-activation. The anti-ferromagnetic Ising model with Metropolis dynamics is adopted and applied to four variations of the large-scale connectivity of the cat thalamocortical system: (a) considering only cortical regions and connections; (b) considering the entire thalamocortical system; (c) the same as in (b) but with the thalamic connections rewired so as to maintain the statistics of node degree and node degree correlations; and (d) as in (b) but with attenuated weights of the connections between cortical and thalamic nodes. A series of interesting findings are obtained, including the identification of specific substructures revealed by correlations between the activity of adjacent regions in case (a) and a pronounced effect of the thalamic connections in splitting the thalamocortical regions into two large groups of nearly homogenous opposite activation (i.e. cortical regions and thalamic nuclei, respectively) in cases (b) and (c). The latter effect is due to the existence of dense connections between cortical and thalamic regions and the lack of interconnectivity between the latter. Another interesting result regarding case (d) is the fact that the pattern of thalamic correlations tended to mirror that of the cortical regions. The possibility to control the level of correlation between the cortical regions by varying the strength of thalamocortical connections is also identified and discussed.

# 1. INTRODUCTION

The mammalian cortex is amongst the most complex structures in nature, not so much because of the large number of constituent neurons, but rather as a consequence of the highly elaborate pattern of interconnectivity between those cells. The computational power of the brain does not result from the speed of its computing elements (neurons) but from their parallel interactions meditated by this pattern of structural connectivity. This pattern is highly adapted to ensure rapid information integration and transmission as a basis for cognition. Not surprisingly, structural connection patterns have been found to be far from trivial, such as would be the case if one cell would connect to all other cells in a given region (e.g. its neighborhood), or in the case of random patterns. Instead, these connections follow asymmetric and diversified patterns of short, medium and long-range connections. It is in great part due to such a complex interconnectivity that human intelligence has emerged. However, this complexity is also one of the main reasons which have severely constrained our efforts in understanding the brain.

Closely related to graphs, the area of complex networks recently emerged as a powerful paradigm which can be used to represent, characterize, model and simulate a broad variety of complex systems including the Internet, protein interactions, social contacts [1] and even partnership in Jazz [2]. Complex networks have already been considered in the context of neuroscience. For example, the nervous system of the nematode *C. elegans* [3] has been extensively analyzed. Mammalian brain connectivity [4-6] was found to be characterized by high degrees of clustering, short path lengths, prevalent connectional reciprocity, specific network motifs and conserved wiring volume (e.g. [7,8]).

Despite these previous studies aiming at the representation and characterization of complex networks, there has been little progress in linking complex network architectures to network dynamics. This exciting area is becoming one of the current focuses of attention in complex network research [9,10]. In large-scale cortical networks, the existence of small-world attributes has been linked with dynamical phenomena such as synchronization (e.g. [11]) and efficient grid computing [12]. How complex brain networks generate coordinated and meta-stable brain states associated with higher brain function is still largely unknown.

A recent study attempted to model the system-wide dynamics of cortex at large spatial scales [13]. After representing the cortical systems as a network, the stochastic matrix defined by a randomly driven Markov chain was obtained. This allowed the steady state of the probability of node occupancy to be determined in terms of the normalized eigenvector associated with the single unit eigenvalue. The relationship between the cortical network and simulated activity was analyzed in terms of the correlation between the node outdegree and the occupancy probabilities. The application of such a methodology revealed the fact that thalamocortical connections (linking individual cortical regions with specific thalamic nuclei) are strengthening this investigated correlation.

The main objective guiding the current work is to address the transient dynamics of cortical regions under conditions promoting the appearance of opposing activity states in pairs of adjacent regions, i.e. those which are directly interconnected. More specifically, we want to investigate how the intrinsic thalamocortical connectivity influences such a kind of dynamics, especially in regard to the correlations of activations between each specific cortical and thalamic region and the remainder of the whole thalamocortical system. The anti-ferromagnetic Ising model under Metropolis dynamics has been adopted. The Ising model, one of the most intensely studied systems in statistical physics, provides an interesting means to complement previous analyses (e.g. [13]) as a result of the consideration of energy-based interactions between pairs of nodes. The choice of anti-ferromagnetic interactions has been motivated by our specific interest in brain dynamics that lead to maximal diversity or variability between the states of individual cortical regions. Such a mode of operation presents a more plausible basis for rich and varied brain states since it avoids more uniform and extreme situations, for example ground states corresponding to widespread activation or silencing of the brain. Such uniform states of cortical activation are more likely to be associated with pathological states, as in uncontrolled epileptic activity or complete lack of activity (brain death), and are thus of less interest as bases of cognition.

In the brain, the influence of an active cortical region on its neighbors usually takes place via long-range excitatory connections, and global states of activation are avoided through regulatory processes involving local inhibitory circuitry. Though it does not represent a direct electrophysiological analogue, the anti-ferromagnetic Ising model does incorporate a possible dynamics for obtaining diversity of neuronal activation while avoiding extreme ground states. Thus, it captures an important characteristic of large-scale brain activity. However, we stress that we do not propose that brain dynamics is directly related to the physics concepts of spin, dipoles, magnetism, or heat, or that dynamics favoring opposing regional activation is the unique, or even typical, mode of operation in the brain. At the same time, the ability of the Ising model to allow the facilitation of opposite states as a consequence of their implied smaller energy, provides a simple and convenient means for making inferences about the types of correlations which can be induced between cortical regions as a consequence of their specific interconnectivity.

Having chosen the type of interaction in the Ising system (i.e. anti-ferromagnetic), a subsequent important decision concerns the computational algorithm for simulating the respective dynamics. Here we adopt the Metropolis methodology [14], representing an accurate approach to simulate Ising systems away from the critical temperature [14]. The basic steps in the Metropolis algorithm, which involves single-spin-flipping, basically emulate the real dynamics of state transitions of an Ising system at equilibrium. The same type of dynamics is assumed for our investigations of the non-equilibrium dynamics. We note that such an approach is inherently arbitrary in the sense of assuming the non-equilibrium dynamics to follow the Metropolis sampling strategy. Nevertheless, such a choice still reflects energy-based interactions between pairs of adjacent regions and therefore provides a reasonable means for investigating dynamics [14] involving interactions between the states of adjacent cortical regions.

The plan of this article is as follows. First, we present the basic materials and methodology, including the origin of the maps of cortical adjacencies, the representation of the cortical structure in terms of a network and the considered measurements of its topology, as well as the basic concepts of the Metropolis method and details of the performed simulations. These concepts are presented in didactic fashion in order to make this text accessible to a wider audience of scientists. Then, we discuss the obtained results, including the identification of the average values of the intensity at each individual region over time and the correlation of these with several topological properties of the cortical network such as degrees, path lengths and clustering coefficients. Several interesting findings have been obtained, including the identification of some specific structures of activation/deactivation patterns in the cortical areas and the pronounced effect of the thalamic connections in splitting the brain dynamics into two largely opposite but stable regions of complementary activation (namely cortex and thalamus). The possibility to modulate the correlation between cortical regions through variations of the intensity of the thalamocortical connections is also suggested and discussed.

## 2. MATERIALS AND METHODS

### 2.1. Cortical Maps

The anatomical data of the cat thalamocortical system used in this paper is the same as that considered in [7,8,13,15] and originally reported in ref. [16] (Figure 1). These publications should be consulted for details regarding the anatomical abbreviations and functional descriptions of specific brain areas. The data itself can be obtained through download from www.indiana.edu/~cortex . Each cortical and thalamic region is treated as a node, so that the large-scale interconnectivity can be represented by the respective adjacency matrix. Of particular interest is the complete absence of interconnections between the thalamic regions. The complete cat thalamocortical network includes 606 directed edges between cortical regions (corticocortical connections) and 1302 directed edges between thalamic and cortical regions (thalamocortical connections). A total of 95 nodes are included in the network, of which the first 53 correspond to cortical regions and the remaining 42 to thalamic regions.

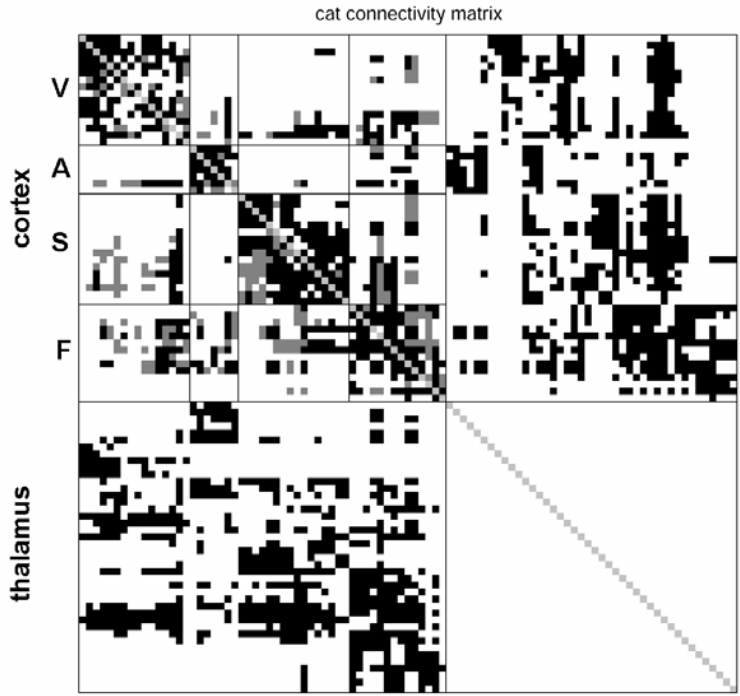

*Figure 1:* Connectivity matrix of cat cortex and thalamus. Cortical regions occupy the first 53 positions of the matrix and thalamic regions occupy the remaining 42 positions. Cortical regions are subdivided into V = "visual", A = "auditory", S = "somatomotor", and F = "frontolimbic" regions. Columns of the matrix contain inputs to a given region, i.e. "targets" are at the top. Gray and black entries mark connections as recorded in the original publication [16]. Connections marked in black correspond to those present in the symmetrized matrix used in this study (see Section 2.2). Gray connections represent unidirectional pathways that have been omitted. No connections are present on the main diagonal.

## 2.2. Complex Networks

Basically, a complex network is a type of graph exhibiting particularly intricate structure of interconnections. As such, complex networks can be properly represented by using the traditional graph notation, involving a set of $N$ nodes $i = 1, 2, \ldots, N$ and $E$ edges represented by tuples $(i, j)$ indicating a connection from node $i$ to node $j$. Such a graph can be conveniently represented in terms of its *adjacency* (also called incidency) matrix $K$, so that the presence of an edge $(i, j)$ implies $K(j, i)=1$. The current article deals on *undirected graphs*, characterized by the presence of bidirectional edges – i.e. $K(j, i)=1$ implies $K(i, j)=1$ and vice-versa.

Because the Ising model assumes reciprocal interactions for every pair of nodes, the directed adjacency matrix obtained for the cortical regions has to be symmetrized as

$$\tilde{K} = \begin{cases} 1 & \text{if } K + K^T \neq 0 \\ 0 & \text{otherwise} \end{cases}$$

This transformation reduces the total of directed edges (symmetric edges counted twice) from 2129 to 1908 (see Figure 1). Since most corticocortical pathways are found to be symmetric, the topological difference between the original and symmetric connection matrices is relatively small. In addition to representing a small reduction of the connectivity of the original cortical structure, the removal of isolated directed edges represents a natural procedure in cases where one is interested in modeling pairwise interactions between nodes.

Given an undirected complex network (or a graph), a series of measurements of its topology and interconnectivity can be obtained. Those of particular interest to the current work are listed and commented in the following:

**(A)** *Node degree*: Given a node $i$, its degree $d(i)$ is equal to the number of edges connected to that node.

**(B)** *Clustering coefficient*: The clustering coefficient reflects how intensely interconnected are the neighbors of a given node $i$,

$$CC(i) = 2 \frac{e(i)}{n(i)[n(i)-1]}$$

where $n(i)$ is the number of neighbors of node $i$ (i.e. those nodes which receive edges emanating from $i$) and $e(i)$ is the number edges among the neighbors of $i$. Note that $0 \leq CC(i) \leq 1$, with unit value indicating maximum interconnection between the neighbors of $i$.

**(C)** *Shortest path between pairs of nodes*: Given two nodes $i$ and $j$, this measurement corresponds to the number of edges along the shortest path connecting those two nodes. Because the considered network is irreducible (any node can be reached from any other node), we do not have to consider infinite shortest paths.

Note that these three measurements apply to a single node (i.e. degree and clustering coefficient) and pairs of nodes (shortest path). In order to obtain characterization of the overall properties of a complex network, we resort to the moments (e.g. average and variance) of such measurements considering all network nodes.

### 2.3. Ising Model

The Ising model was introduced by Wilhelm Lenz as a suggestion for the PhD of his student Ernst Ising in the 1920s [17]. It is a model of a magnet, where each node corresponds to a magnetic dipole and the links between pairs of adjacent nodes express the pairwise, reciprocal, magnetic interactions. The Hamiltonian for such a system underlined by an orthogonal lattice containing $N$ dipoles is given as

$$E = -J \sum_{i, j \in Q(i)} s_i s_j - B \sum_i s_i \qquad (1)$$

where $s_i$ and $s_j$ are the states of the dipoles $i$ and $j$ (such values are limited to +1 and -1) $J$ is the interaction parameter, $Q(i)$ is the set containing the neighbors of $i$ in the lattice and $B$ is an external magnetic field interacting with the system. As in many Ising simulations, $B$ is assumed to be null in the present work.

The mean magnetization of the system at any time instant is given as

$$\langle M \rangle = \frac{1}{N} \sum_i s_i$$

At equilibrium, the occupation probability of each overall state $\alpha$ of the system, defined by the individual states of each dipole, follows the Boltzmann distribution, i.e.

$$p_\alpha = \frac{1}{Z} e^{-E_\alpha/(kT)} \qquad (2)$$

where $Z$ is the partition function, $k$ is the Boltzmann constant, $T$ is the temperature of the reservoir in which the Ising system is immersed, and $E_\alpha$ is the energy of state $\alpha$ (given by Equation 1). For simplicity, we henceforth assume $\beta = 1/(kT)$. By supplying or absorbing energy, the reservoir allows the spins of the dipoles to change, producing a dynamical evolution of the states.

Two basic types of interactions are typically considered: (i) *ferromagnetic*, characterized $J = -1$; and (ii) *anti-ferromagnetic*, characterized $J = 1$. It can be verified from Equations (1) and (2) that, for small temperatures, the ferromagnetic mode of interaction will cause the spins of the dipoles to align in pairwise fashion. Contrariwise, in the anti-ferromagnetic model, the spins of neighboring dipoles will tend to become opposite. As justified in the Introduction, in this article we concentrate our attention in anti-ferromagnetic Ising model.

While all the above concepts assume the Ising system to have the dipoles distributed along an orthogonal lattice, it is possible to immediately adapt such a system to a graph/network. This can be done by associating each dipole to a network node, while the neighborhoods in Equation 1 are defined by the adjacencies between nodes. However, while all dipoles in the traditional Ising system have identical connectivity, this is no longer true for a graph, where distinct dipoles may exhibit completely different interactions. For instance, the flipping of the spin of a hub (i.e. a node with high degree) will tend to produce a variation of energy much larger than for a node with low degree.

The clustering coefficient is also closely related to the possible energy changes and state configurations in such graph-underlined Ising models.

Because of the immense number of possible states which a reasonable sized Ising system can exhibit (the number of such states is equal to $2^N$), it is virtually impossible to obtain the equilibrium distribution while considering all states. A *sampling strategy* is required for that. The current work adopts the Metropolis algorithm for sampling the Ising states. Starting at a given initial state $\alpha_0$ with respective energy $E(\alpha_0)$, subsequent putative states are produced by changing the spin of one of the dipoles chosen randomly among the $N$ total dipoles in the system. The energy of this state, $E(\alpha)$, is calculated so that the increase of energy while moving from state $\alpha_0$ to state $\alpha$ corresponds to $\Delta E = E(\alpha) - E(\alpha_0)$. The new state $\alpha$ is accepted whenever $\Delta E < 0$ or with probability $p(\Delta E) = \exp\{-\beta \Delta E\}$ otherwise (i.e. $\Delta E \geq 0$).

While all the equations and methods above assume the Ising system to be at equilibrium, it is also possible (and interesting) to consider transient behavior of the individual states while the system evolves out of equilibrium. The main problem with such a kind of investigation is that the dynamics out of equilibrium is usually not known, so that one has to impose some arbitrary dynamics. Still, interesting insights can be inferred in this way [14]. As we are interested in the transient behavior of the system at a temperature relatively far from the critical one, it is reasonable to adopt the Metropolis algorithm for generating the dynamics. All simulations in this work consider $T = 0.4T_C = 0.8J / \log(1 + \sqrt{2})$. A total of 200 realizations are considered for each of the investigated cortical architectures, and in each case the system is allowed to evolve along 1000 spin flippings. The initial state corresponds to taking all individual states equal to one and changing the sign of a total of $0.4N$ states uniformly random fashion. The characterization of the individual dynamics of each of the $N$ states is accomplished by considering the pairwise correlations between every dipole in the system along the 1000 time steps, which is conveniently represented as a $N \times N$ correlation matrix $CC$ estimated for each simulation. The overall properties of all realizations performed for each cortical architecture are expressed in terms of the mean and standard deviation of $CC$.

Previous investigations involving complex networks and Ising models include the consideration of community finding [18], the modeling of opinion formation [19], interactive statistical mechanics algorithms [20], and the investigation of correlations in terms of successive neighborhoods [21].

## 3. RESULTS AND DISCUSSION

In this work, we consider the transient dynamics of anti-ferromagnetic Ising models with connectivity corresponding to the three following situations: (i) the network with only cortical regions and connections; (ii) such a network plus the thalamic regions and connections; (iii) the network in (ii) but with the thalamic connections scrambled so as to maintain the node degree and node degree correlations (e.g. [15]); and (iv) the same as (ii) but with attenuated thalamocortical connections. Each of the four considered interconnecting architectures are addressed in the following:

### *3.1. Cortical Regions and Connections*

For this case we consider only the 53 cortical nodes and all their corticocortical interconnections. The pattern of mean correlations between each pair of dipoles estimated over 1000 spin-flippings for all 200 realizations is shown in Figures 2A. Clearly, we obtain a distinct and structured pattern of correlations that is induced by the specific pattern of corticocortical interconnectivity. Correlation values are spanning virtually the entire interval of [–1 1] and most correlations are distributed around zero (Fig. 3A). The pattern of correlations does not trivially mirror the underlying pattern of structural connections. As shown in Fig. 1, the original connection matrix is ordered into groupings of sensory and motor regions. While ferromagnetic coupling or standard excitatory interconnectivity tends to produce activations that directly reflect this group structure, with regions within groups more strongly correlated with each other, anti-ferromagnetic coupling as employed in this study produces a more diverse and less homogeneous pattern. For example, subsets of visual regions coupled via shared reciprocal pathways exhibit differentiated patterns of positive and negative correlations (circled in Fig. 2). In order to investigate the nature of these heterogeneous patterns we examined the relation of sign and magnitude of the observed correlations and the distance *d(i,j)* between each pair of regions (Fig. 4). We found that pairs of regions that are linked by a direct connection (*d(i,j)* = 1) tended to be anti-correlated, while regions at a distance of *d(i,j)* = 2 tended to be correlated.

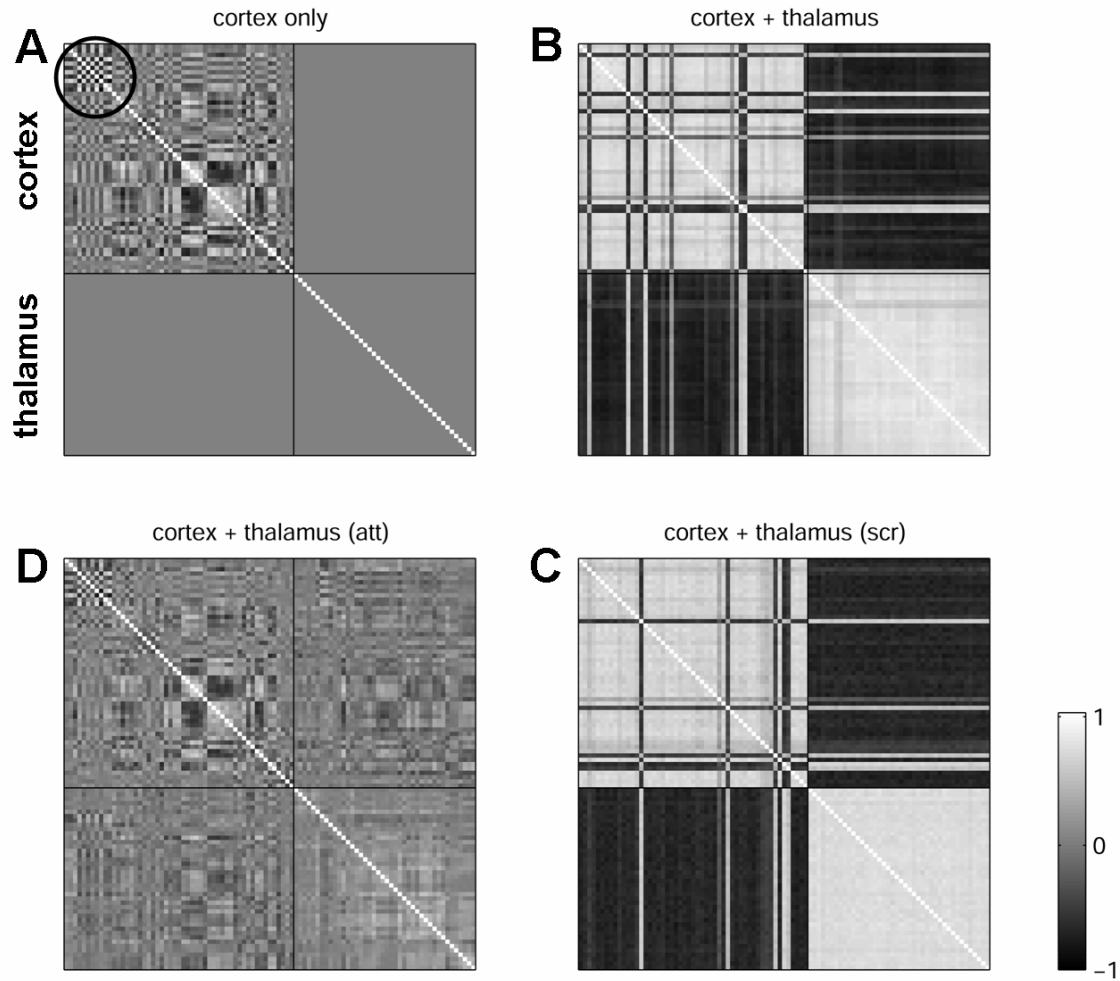

*Figure 2:* *Means of correlation patterns for four cases examined in this study. All plots show correlations in the interval [−1 1] (see scale bar at right), with 53 cortical regions arranged in the upper right of the matrix and 42 thalamic regions arranged in the lower left of the matrix. Thalamocortical correlations are shown in the two off-diagonal rectangles. (A) Cortical regions only. Circle highlights pattern in visual areas. (B) Cortical and thalamic regions with the original anatomy preserved. (C) Cortical and thalamic regions with the thalamocortical connections randomized ("scrambled"). (D) Cortical and thalamic regions with the thalamocortical connections attenuated.*

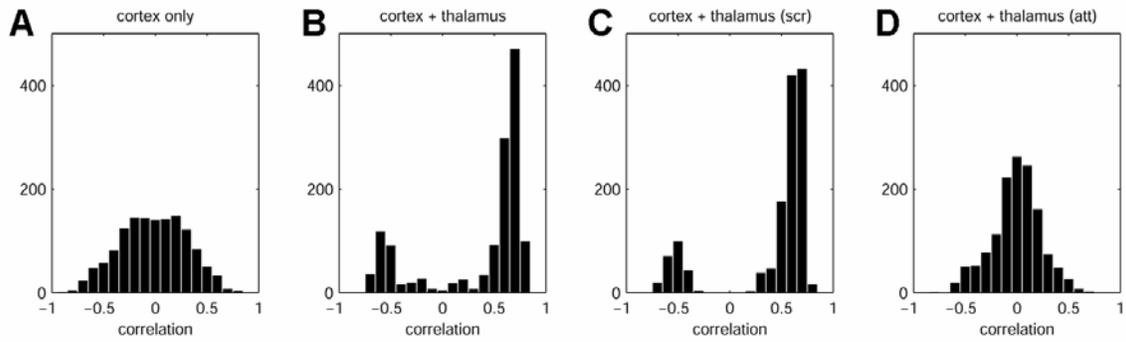

*Figure 3:* Histogram distributions of corticocortical correlations (a total of 1378 values from the upper triagonal of the corresponding cortical correlation matrices shown in Figure 2).

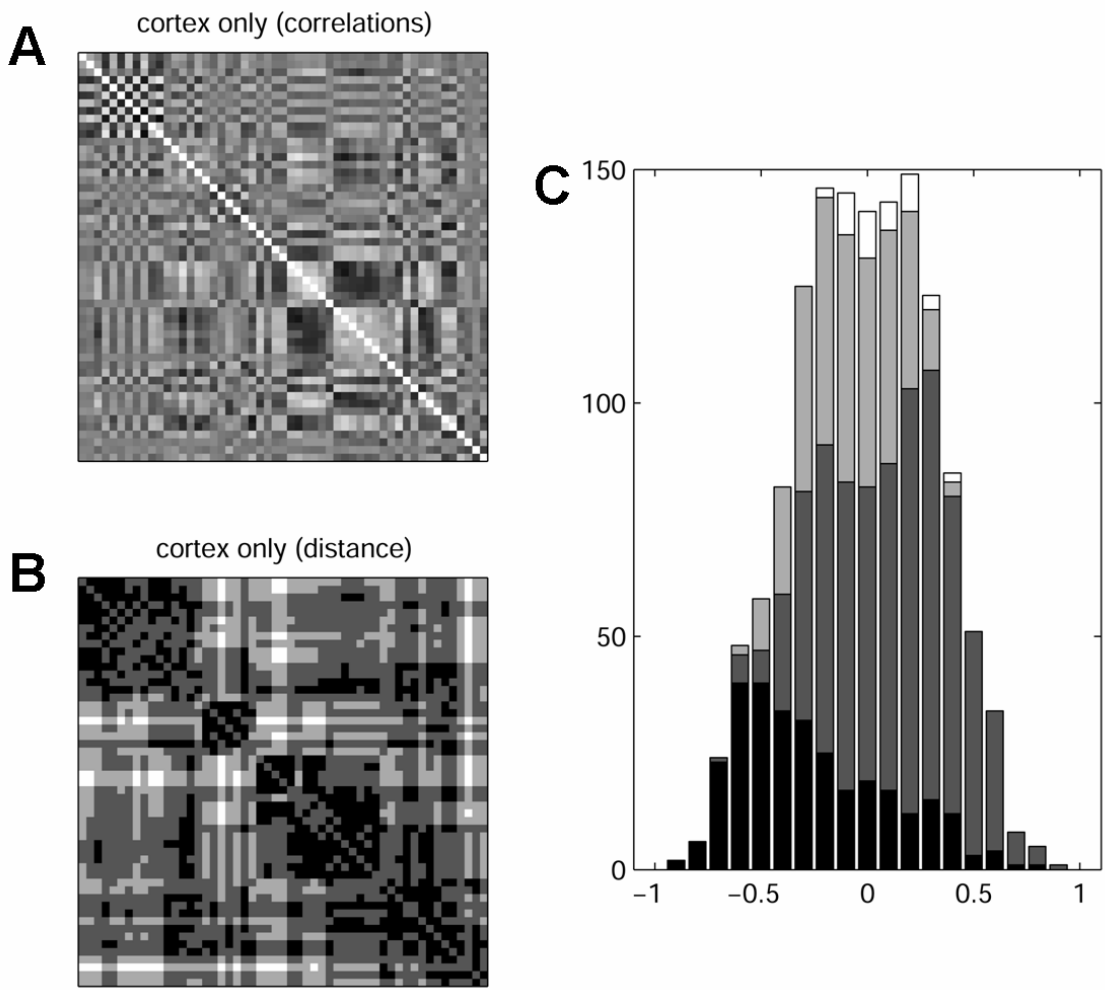

*Figure 4:* Comparison of correlation patterns and distance matrix for correlations between cortical regions in isolation. (A) Mean correlation matrix for 53 cortical regions (compare Fig. 2A). Grayscale for correlations is the same as that for Figure 2. (B) Distance matrix for 53 cortical regions obtained from the symmetrized connection matrix (Figure 1; see Section 2.2). Black corresponds to a distance of 1, white corresponds to a distance of 4, with two intermediate gray levels coding for distances of 2 (dark gray) and 3 (light gray). (C) Distribution of correlations shown as a stacked histogram plot. The outer contour of the plot is identical to that in Figure 2A. The gray scale is identical to the one used for the distance matrix. Most correlations between vertices at a distance of 1 (black bars) are negative. Most correlations between vertices at a distance of 2 (dark gray bars) are positive.

### 3.2. Cortical plus Thalamic Regions and Connections

The mean pattern of correlations obtained for the Ising simulation considering all original cortical and thalamic nodes and their connections are given in Figure 2B. Compared to the results shown in Figure 2A, the pattern of spin correlations has been radically altered by the added thalamocortical connections. This can be clearly seen by comparing the upper left 53 × 53 portion of the matrix in Figure 2A with the corresponding portion of the matrix in Figure 2B. The histogram distribution of correlations was changed from a symmetric unimodal distribution centered around zero (Fig. 3A) to an asymmetric bimodal distribution, with most correlations on the positive side (Fig. 3B).

This distribution captured the fact that the majority of cortical regions are now very strongly correlated, with the exception of a few areas (black stripes of anti-correlations at the upper left rectangle in Figure 2B). The incorporation of the thalamic connections resulted in a major reorganization of the whole structure into two groupings of nearly homogeneous and opposing activation, corresponding to the cortical and thalamic regions. Each of these two sets of regions is characterized by internal positive correlation while being anti-correlated to one another. It should be noted that such overall organization corresponds, in the mean, to active cortical regions counterbalanced by inactive thalamic regions or vice-versa. Therefore, overall stability is achieved because the positive energies implied by the correlations among the cortical regions (remember the lack of connections between thalamic regions) are counterbalanced by the negative energies between cortical and thalamic regions.

The explanation of this critical phenomenon is related to the intense interconnectivity between cortical and thalamic regions and the lack of interconnectivity between the latter. The diagram in Figure 4 illustrates in a rather simplified fashion the key elements involved in the partition phenomenon. First, cortical regions are less interconnected (606 directed edges, connection density = 21.9%) than cortical-thalamic regions (1302 directed edges, connection density = 28.6%). At the same time, the thalamic regions are not directly interconnected. Consider any two cortical regions *A* and *C*, both of which are connected to a thalamic region *B*, as in Figure 5. In order to allow lower overall energy, the state of *B* should be the opposite of that in region *A* (e.g. +1 and -1, as in the figure).

The same reasoning applies to the states in nodes *B* and *C* (e.g., -1 and +1, respectively). Therefore, though opposite states arise in *A*,*C* and *B*, the states in *A* and *C* tend to be the same as a consequence of the transitivity of the opposite states in *A*/*B* and *B*/*C*. Therefore, a 'virtual' ferromagnetic edge is established between *A* and *C* as a consequence of the transitivity of the two successive anti-ferromagnetic connections, favoring the uniformity of the states between the cortical regions. Because there are more thalamocortical connections (favoring virtual ferromagnetic organization between cortical regions) than the anti-ferromagnetic connections within the cortex, the effect of the former tends to prevail over the latter, yielding positive correlations between the cortical activities.

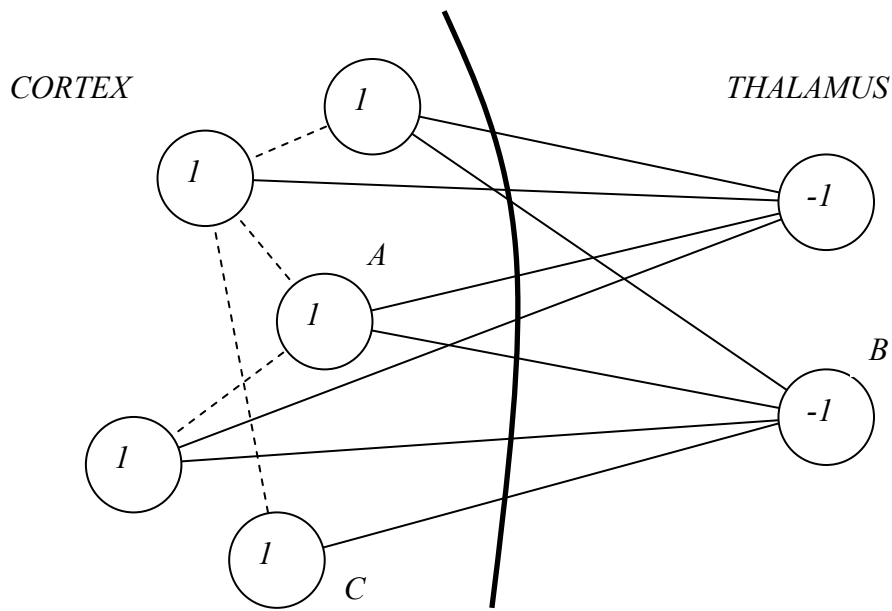

*Figure 5: The lack of interconnectivity between thalamic regions together with the high density of interconnections between cortex and thalamus result in stable patterns of overall activation involving uniform configurations of states at cortical and thalamic regions. Observe that the "V" connections implemented by the thalamic regions tend to favor positive correlation between cortical regions. For instance, if a cortical region A is connected to another cortical region C through a thalamic region B, A will tend to be positively correlated (aligned activities) with C as a consequence of the transitive anti-correlations between A/B and B/C.*

### 3.3. Cortical plus Scrambled Thalamic Regions and Connections

In this section we consider the original cortical regions and their corticocortical interconnections, but the thalamocortical connections have been randomized (or "scrambled") as described at the beginning of this section. The mean pattern of the obtained spin correlations are shown in Figure 1C.  The investigation of this type of architecture has been motivated by the need to infer to which extent the original thalamocortical connections are specific and purposive.  This is particularly interesting because we found, in a previous investigation [13], that the thalamocortical connections were effectively organized so as to enhance the correlation between the average occupation of each node in a Markov random walk model (analogous to the node's activation level) and its outdegree.  Two important conclusions can be drawn from the results shown in Figure 2C: (i) the overall pattern of spin correlations observed for the scrambled thalamocortical links (i.e. the separation of thalamic from cortical regions) is similar to that obtained for the original thalamocortical connections (see Figure 2B); at the same time (ii) the fine structure of the correlations among both the cortical and thalamic regions is clearly different from that shown in Figure 2B.   The fact that the overall pattern of correlations appeared very similar to that obtained for the original thalamocortical systems (i.e. partitioned into two nearly uniform regions, cortex and thalamus) corroborates the explanation provided in the previous Section.  In other words, the same partitioning effect of activities has been obtained by considering a completely different interconnectivity between cortical and thalamic regions, except for keeping the number and degree correlation between thalamic regions, as well as their lack of interconnectivity, which were essential for reproducing the partition phenomenon.

### *3.4. Cortical and Attenuated Thalamic Regions and Connections*

Though interesting, the phenomenon of bipartition of the correlations of the whole thalamocortical system into two regions of nearly uniform activation actually seemed to contradict the initial rationale for adopting the anti-ferromagnetic model as a means for promoting diversified activity among adjacent cortical regions.  The results shown in Figure 2B and 2C motivated a re-assessment of the data and model.  It became apparent that the assignment of weights to thalamocortical connections which were equal to the weights of the corticocortical interconnections had a major effect on these results.  Indeed, the assignment of equal weights assumes that the influence of a small thalamic nucleus (perhaps containing only $10^4$ or $10^5$ cells), through its total neuronal activity, would be equal to that of a large cortical region (comprising on the order of $10^7$ neurons).  Arguably, a more reasonable configuration would consider the intensity of the connections between cortical and thalamic regions to be a fraction *f* of the weight of the cortical interconnections.  Additional simulations were performed in order to investigate this possibility.  Figure 2D shows the mean pattern of correlations obtained for the thalamocortical activities considering *f* = 0.2.  Interestingly, not only were the pattern and distribution (Fig. 3D) of correlations and anti-correlations obtained for the isolated cortical regions (Section 3.1) recovered, but significant and differentiated correlations were also observed between cortical regions and thalamic regions.  Also of interest is the fact that correlations between the thalamic regions (lower right-hand square in Figure 2D) now appeared structured and diversified.  Additional simulations (not shown in this article) indicated that it is possible to implement progressive modulatory activity of the thalamic structures over the correlations between cortical regions by varying the intensity of the connections between the cortical and thalamic regions.

Another interesting feature in Figure 2D concerns the fact that the pattern of mean correlations between thalamic regions obtained by attenuating the thalamocortical connections is similar to the pattern of mean correlations between the original cortical regions in Figure 2A. In order to investigate the statistical plausibility of this hypothesis, the 42 × 42 matrix of thalamic correlations in Figure 2D was interpolated into a 53 × 53 matrix by using the Fourier transform (the thalamic matrix is Fourier transformed and the missing elements are padded with zero, so that the inverse Fourier transform corresponds to the interpolation) and the distance $D_0$ between such an interpolated matrix and the original mean cortical correlation matrix was calculated. The Euclidean distances $D$ between the original mean cortical correlations and a total of 500 randomly scrambled (obtained through column and row swapping) versions of the interpolated matrix were also determined and shown in the histogram in Figure 6. The distance $D_0$ is also shown as the asterisk in this figure, suggesting that the thalamic correlations obtained by using attenuated interconnections are significantly more similar to the original cortical correlations than the ensemble of randomly scrambled matrices. It should be noticed that a substantially greater disparity would have been revealed if completely random matrices were considered instead of the scrambled versions of the thalamic correlations.

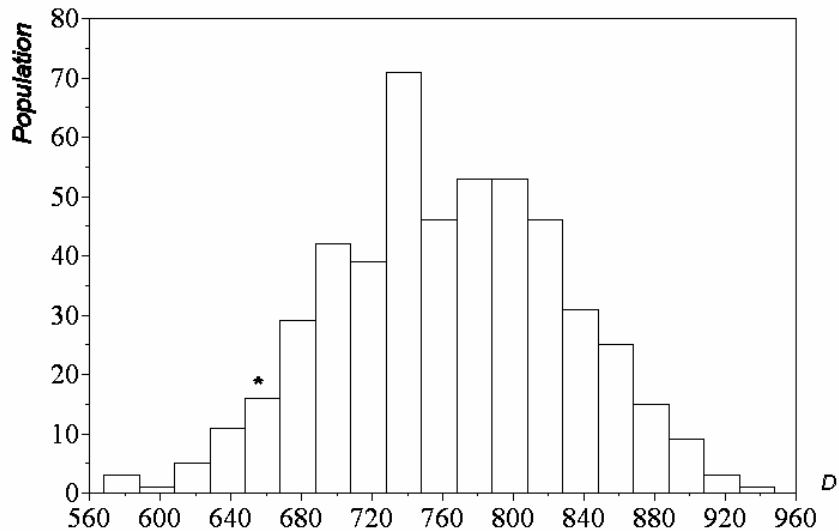

*Figure 6:* *The distribution of the Euclidean distances D between the original cortical correlations (i.e. the matrix in Figure 2A) and 500 scrambled versions of the interpolated mean thalamic correlations obtained for attenuated connections (i.e. the lower right-hand matrix in Figure 2D). The asterisk indicates the difference obtained without scrambling.*

## 4. RELATING CONNECTIVITY AND CORRELATIONS

While the consideration of the correlations between thalamocortical activations along non-equilibrium evolution led to a series of interesting results, it is also useful to investigate how the Ising correlations are possibly related to the topological properties of the cortical network. One such relationship is illustrated in Figure 4, showing a relationship between topological distance and correlation values. To further investigate possible relations between correlations and topological indices, we defined the total correlation of each node $i$ as corresponding to the sum of the correlations such a node exhibits with all the other cortical nodes. The scatter plots obtained by considering such a measurements against the three topological features presented in Section 2.2 are shown in Figure 7A,B,C, respectively. Except for a small negative correlation observed between the total correlation and node degree in Figure 7A, no other significant correlation is observed for the two other relationships. Also, there was no significant difference between two major groupings of cortical regions – i.e. posterior (visual plus auditory) and anterior (somatomotor plus frontolimbic) – for any of the three cases (data not shown). These results are a consequence of the fact that the Ising energy configurations, involving transient effects such as those discussed in section 3.3, do not strongly depend on local topological measures such as node degree and clustering coefficient, but mainly on longer-range topological properties of the network. Because the widely used local topological measures are generally unable to reflect important aspects of the considered Ising dynamics, the interesting effects identified in the previous sections would be likely be missed were it not for the explicit consideration of the non-equilibrium dynamics of the Ising model.

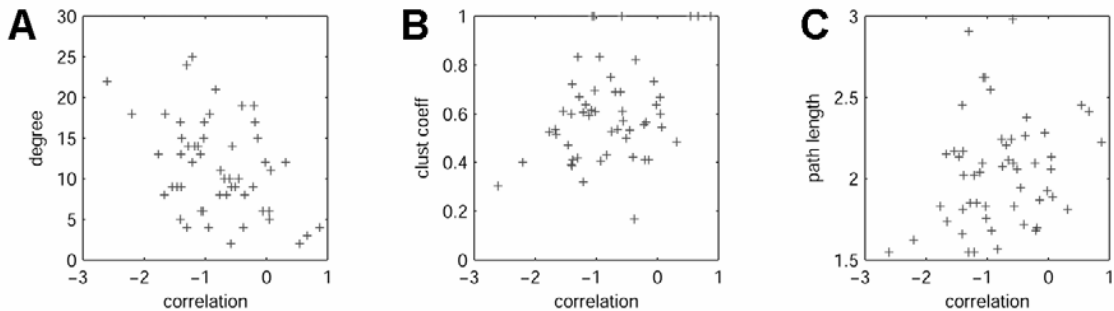

*Figure 7:* *Scatter plots of degree (A), clustering coefficient (B) and path length (C) versus total correlation (all measures per each cortical node).*

## 5. CONCLUDING REMARKS

This article has investigated how the interconnectivity of the thalamocortical system of the cat brain constrains and defines the distribution of correlations between cortical activations under an imposed dynamics promoting opposing activations between adjacent regions. The anti-ferromagnetic Ising model under the Metropolis dynamics was adopted as the means to quantify the correlations between regions during non-equilibrium evolution. Although the biological plausibility of this model possibly extends no further than implementing the tendency of opposite activities among adjacent regions, it has provided a number of interesting findings. These include the specific patterns of correlation between cortical regions and the fact that the thalamic connections can lead to the partitioning of the system into two groupings of nearly uniform positive correlations, corresponding to the cortical and thalamic regions, an effect which has been explained as a consequence of the high density of interconnections between the cortex and thalamus and the lack of interconnectivity between thalamic nodes, as corroborated by the consideration of a rewired system maintaining the node degree and node degree correlations between the thalamic regions.  It has been also shown that this effect can be modulated by reducing the weight of the connections between cortical and thalamic regions, suggesting an interesting putative regulatory role for the thalamus. The correlations obtained for such an attenuated system also exhibited distinctive connectivity patterns between the cortical regions, including the interesting fact that the correlation between thalamic regions exhibited significant similarity to the pattern of original cortical correlations. Finally, we have also shown a general absence of correlations between the spin total correlations and traditional topological features of the cortical network, indicating that the specific attributes of the Ising dynamics in the thalamocortical system are regulated by more global interconnectivity measures. Further related investigations could consider the interconnectivity in cortical systems of other species, as well as the use of the Potts model [14] as a means to provide representation of more gradual cortical activities.

## ACKNOWLEDGMENTS

Luciano da F. Costa thanks CNPq (308231/03-1) for sponsorship.  Olaf Sporns was supported by the James S. McDonnell Foundation.